\documentclass[12pt,linenumbers]{article}

\usepackage[a4paper,left=30mm,right=20mm,top=20mm,bottom=20mm]{geometry}
\usepackage{graphics}
\usepackage{amsmath}
\usepackage{epsfig}
\usepackage{cite}
\usepackage{lineno}
\usepackage[varg]{txfonts}
\newcommand{\be}{\begin{eqnarray}}
\newcommand{\ee}{\end{eqnarray}}
\title{Self-consistent analysis  of hadron production 
in $pp$ and $AA$ collisions at mid-rapidity}

\author{G.I. Lykasov, A.I. Malakhov} 
\begin{document}
\date{}
\maketitle

\begin{center}

{Joint Institute for Nuclear Research, Dubna 141980, Moscow region, Russia

\vspace{0.5cm}

%artemenkov.denis@gmail.com\\
lykasov@jinr.ru\\
malakhov@lhe.jinr.ru\\
%jerus@jinr.ru
}

\end{center}

\vspace{0.5cm}

\begin{center}

{\bf Abstract }

\end{center} 

%\indent
%{\bf
The self-consistent approach based on similarity of inclusive spectra of hadrons produced in $pp$
and $AA$ collisions is reviewed. 
This approach allows us to describe rather well the
ratio of proton to anti-proton yields in $AA$ collisions as a function of the initial energy at a wide 
range from a few GeV to a few TeV. 
We suggest its modification due to the quark-gluon dynamics to describe the inclusive spectra of hadrons 
produced in $pp$ collision as a function of the transverse momentum $p_t$ at mid-rapidity. 
The extension of this approach to analyze the pion $p_t$-spectra produced in
$AA$ collision at high and middle energies and mid-rapidity is given. The satisfactory description 
of experimental data on these spectra in $pp$ and $AA$ collisions within the offered 
approach is shown.
%}
     
\vspace{1.0cm}

\noindent
%PACS number(s): 12.15.Ji, 12.38.Bx, 13.85.Qk

%\newpage
\indent
%%%%%%%%%%%%%%%%%%%%%%%%%%%%%%%%%%%%%%%%%%%%%%%%%%%%%%%%%%%%%%%%%%%%%%%%%%%%%%%%%%%%%%%%%%%%%%%%%%%%%%

\section{Introduction}
\label{intro}
The similarity principle, for example, in physics is well known. Its application to the particle production in hadron-hadron collision
 has been proposed
several decades ago by E.Fermi \cite{Fermi:1950}, I.Pomeranchuk \cite{Pomeran:1951}, L.D.Landau \cite{Landau:1953} and 
R.Hagedorn 
\cite{Hagedorn:1965,PBM:2013}. 
It was noticed that the transverse momentum  spectra of the particles produced in these collisions had a universal form like
 $\rho_h\sim\exp(-m_{ht}/T)$, at least,
at not large values of $m_{ht}$, where $m_{ht}$
 is the transverse mass of the produced hadron $h$ and $T$ is a constant dependent only of the type of final
hadrons. 
There are many statistical models applied successfully a similar form to describe hadronic 
yields produced in heavy-ion 
collisions (see, for example, \cite{Chatt:2015,Bugaev:2018,Br-Munz}
and references therein).  
Actually, the parameter $T$ is nothing to do the average transverse mass $\langle m_{ht} \rangle$. However, according 
to many experimental data, the average transverse momentum $p_{ht}$ of the produced hadron or $m_{ht}=\sqrt{p^2_{ht}+m^2_{h}}$ depends on the initial 
energy $\sqrt{s}$, especially at low $\sqrt{s}$. Here $s$ is the energy squared in the c.m.s. of colliding particles and $m_{h}$ is the mass of the produced hadron. 

Almost all theoretical approaches operate with the relativistic invariant Mandelstam variables $s,t,u$
to analyze the hadron inclusive spectra in the  mid-rapidity region. As usual, the spectra are presented in the factorized forms of 
two functions dependent of $t$ or $p_t^2$ and $s$.
However, there is another approach to analyze multiple hadron production in $pp$ and $AA$ collisions at high energies,
which operates by using four velocities of the initial and final particles \cite{c3}. It is the so called ``the self-similarity 
approach'', which demonstrates the similarity of inclusive spectra of hadrons produced in $pp$ and $AA$ collisions, as a function
of similarity parameter $\Pi$. The approach of studying relativistic nuclear interactions in the four 
velocity space proved to be very fruitful \cite{c4}. 
In fact, this approach is valid not in the complete kinematical region. That will be discussed in our paper.
The hadron inclusive spectra obtained within this approach are presented as a function of the relativistic invariant similarity parameter, 
which can be related to variables $t$ and $s$, as it is shown in \cite{ALM:2015}. The general form of such spectrum is not factorized over
$t$ and $s$. 
In this paper we present further development of this approach and extend it to the hadron production in $AA$ collisions at the center 
rapidity region. 
\section{The parameter or function of self-similarity $\Pi$.}
\label{sec:1}

  The description of multi particle states of the relativistic nuclear physics in terms of macroscopic variables such as
temperature, pressure, density, entropy, contradicts to an important principle, which is emphasized by
Heisenberg \cite{a1}: physical laws and the approval must be expressed only
within the observed values.
In the study of collisions of the relativistic nuclei most of these
macroscopic variables have not been observed.
In this work we have used the approach based on the law of similarity
which is applied in relativistic
nuclear physics but not based on Lagrange method. This allows one to design
the solution based on the above
principles. In hydraulics, as it is well known, methods of the theory of
dimensionality and similarity are widely used.
In fact, the invariant relations between the measured parameters of the
problem are determined. In hydrodynamics
the methods of similarity are widely applied they often are the only mean
of the equation analysis.

The versatility of the methods of similarity for the theory and experiment
is not accidental. The fact is that similarity
transformations define the invariant relationships which characterize the
structure of all the laws of nature, including
the laws of relativistic nuclear physics \cite{a2}.
In analogy with the geometrical similarity for physical phenomena we have
used invariant dimensionless combinations
(similarity parameters), composed from the dimension values defining the
task.

     For example, when planning large expensive hydraulic structures it is
necessary to carry out physical modeling.
Geometrically, the body of model is made similarly to the nature-body. For
the steady motion of a viscous incompressible
fluid flowing over the body, the directions of the velocities in the model
and nature are the same.

     As the main parameters of the problem we have taken the following:
$l$ -  the characteristic size of the body model, $l^0$ is the size of 
the nature body, $l^0 / l$ is the coefficient of geometric similarity, 
and $U$ is the velocity of the impinging flow,
$\mu$ is the viscosity of the fluid, $\rho$ is the fluid density.

     These parameters define the system of units: $l$ - length, M -  mass, 
T - time, and have the following dimensions:
$$[l] = L,  [U] = L \cdot T^{-1}, [ \mu ] = M \cdot L^{-1} \cdot T,
  [\rho] = M \cdot L^{-3}.$$

     From the defining parameters we can construct only one dynamic
similarity parameter (a dimensionless combination, independent of the
choice of measuring units):

\centerline{$ \Pi = \rho U l / \mu = $Re. }

     This invariant is called the Reynolds number. To provide the
similarity, it is required to have equality of this parameter for the model
and nature. If the distance {\bfseries r} is measured in units of $l$ and the 
velocity {\bfseries V} is measured in units of $U$, the solutions of the 
hydrodynamic equations for the velocity distributions will have the 
following form:

\centerline{ {\bfseries V}$/U$ = {\bfseries f}({\bfseries r}/$l$, Re).}
  
   From this expression it follows that the fields of the flow velocities
around geometrically similar bodies are described by only one function,
depending on {\bfseries r}$/l$ if the Reynolds numbers for these flows are the same.
%%%%%%%%%%%%%%%%%%%%%%%%%%%%%%%%%%%%%%%%%%%%%%%%%%%%%%%%%

Within the self-similarity approach \cite{c3,c4} the predictions on 
the ratios of the particles produced in $AA$ collisions at high energies were given in \cite{a2,c5}.
Let us briefly present here the main idea of this study. Consider, for example, 
the production of hadrons $1,2$, etc. in the collision of nucleus $I$ with nucleus II:
\begin{equation}
I  +   I\!I  \to  1  +  2  + \ldots                                           
\label{eq:n1}                                                                
\end{equation}                           
    
    According to this assumption more than one nucleon in the nucleus I can participate in 
the interaction (\ref{eq:n1}). The value of $N_I$ is the effective 
number of nucleons inside the nucleus $I$, participating in the interaction 
which is called ``the cumulative number''. 
Its values lie in the region 
of $0 \le N_I \le A_I$ ($A_I$ - atomic number of nucleus I). The cumulative area complies with $N_I > 1$. Of course, 
the same situation will take place for the nucleus $I\!I$, and it is possible to introduce the 
cumulative number of $N_{I\!I}$.

    For reaction (\ref{eq:n1}) with the production of the inclusive particle 1
we can write the conservation law of four-\\
momentum in the following form: 
$${(N_IP_I + N_{I\!I}P_{I\!I} - p_1)}^2 = $$
\begin{equation}
{(N_Im_0 + N_{I\!I} m_0 + M)}^2 ,
\label{eq:n2}
\end{equation}

\noindent where $N_I$ and $N_{I\!I}$ 
the number of nucleons involved in the interaction; 
$P_I$, $P_{I\!I}$ , $p_1$
are four momenta of the nuclei $I$ and ${I\!I}$ and particle $1$, 
respectively; $m_0$
is the mass of the nucleon; $M$ is the mass of the particle 
providing the conservation of the baryon number, 
strangeness, and other quantum numbers.

    In \cite{a2} the parameter of self-similarity is introduced, which allows one to 
describe the differential cross section of the yield of a large class of 
particles in relativistic nuclear collisions:

\begin{equation} 
\Pi=\min[\frac{1}{2} [(u_I N_I + u_{I\!I} N_{I\!I})^2]^{1/2} ,
\label{eq:n3} 
\end{equation}

where $u_I$ and $u_{I\!I}$  are four velocities of the nuclei  $I$ and 
${I\!I}$. The values $N_I$ and $N_{II}$ will be measurable, if we accept the hypothesis of minimum
 mass $m_{0}^{2}(u_1N_1 +u_2N_2)^2$ and consider the conservation law of 4-momentum. Thus,
 the procedure to determine $N_I$ and $N_{II}$, and, hence, $\Pi$, is the determination of the
 minimum of $\Pi$ on the basis of the conservation law of energy-momentum.

Then, it was suggested \cite{a2,c5} that the inclusive spectrum of the produced particle $1$ in $AA$ collision
can be presented as the universal function dependent of the self-similarity parameter $\Pi$, which was chosen,
for example, as the exponential function:

$$ E d^3 \sigma/dp^3 = $$
\begin{equation}
C_1 A_I^{\alpha(N_I)} \cdot 
A_{I\!I}^{\alpha(N_{I\!I})} \cdot \exp(-\Pi/C_2),
\label{eq:n4} 
\end{equation}

where $\alpha(N_I)=1/3 + N_I/3$, \\
$\alpha(N_{I\!I})=1/3 + N_{I\!I}/3$, \\
$C_1=1.9 \cdot 10^4 $ mb $\cdot$ GeV$^{-2}$ $\cdot$ c$^3$ $\cdot$ st$^{-1}$ 
and \\
$C_2 = 0.125 \pm 0.002$.

%%%%%%%%%%%%%%%%%%%%%%%%%%%%%%%%%%%%%%%%%%%%%%%%%%%%%%%%%
%%%%%%%%%%%%%%%%%%%%%%%%%%%%%%%%%%%%%%%%%%%%%%%%%%%%%%%%%%%%%%%%%%%%%%%%%%%%%%
%\subsection{Self-similarity function $\Pi$ in the central rapidity region}
%\label{se3}

\subsection{Relation of self-similarity function  $\Pi$ to the Mandelstam variables $s,t,u$}
%Further development of Baldin's  approach.}
\label{se4}

As it is mentioned above, the exponential form for the hadron inclusive spectrum given by Eq.(\ref{eq:n4})
was chosen as an example and using it we can satisfactorily describe the ratio of total yields of 
antiprotons to the protons produced in heavy nucleus-nucleus collisions. Unfortunately, this simple 
form Eq.(\ref{eq:n4}) contradicts to the LHC data on the inclusive spectra of hadrons produced in the 
central $pp$ collisions, as shown in \cite{c12}. Therefore, we use the results of \cite{c12} 
to present the inclusive relativistic invariant hadron spectrum at the mid-rapidity region and at 
not large hadron transverse momenta $p_{t}$ in a more complicated form, which 
consists of two parts. The first one is due to the contribution of quarks, which was obtained within
the QGSM (Quark-Gluon String Model) \cite{ABK:1982,ABK:1999} using the AGK (Abramovsky, Gribov, Kanchelli) 
cancellation \cite{c10} of $n$-pomeron exchanges for inclusive hadron spectra in the mid-rapidity region.
It is written in the following form \cite{c12}: 

$$E(d^3 \sigma/d^3p)_q =
\phi_q(y=0,p_t) \cdot \sum_{n=1}^{\infty}[n \sigma_n(s)] =$$ 
\begin{equation}
\phi_q(y=0, p_t) g(s/s_0)^{\Delta}~,
\label{eq:n14}
\end{equation}
where $\sigma_n(s)$ is the cross-section to produce the $n$-pomeron chain (or 2n quark-antiquark strings);
$g=21$ mb - constant, which is calculated within 
the "quasi-eikonal" approximation \cite{c11};
$s_0 =$ 1~GeV$^2$; $\Delta = [\alpha_p(0)-1] \sim 0.12$, where $\alpha_p(0)$ is the sub critical 
Pomeron intercept \cite{ABK:1982,ABK:1999,c11}.                     

The second part of the hadron inclusive spectrum at the mid-rapidity region was introduced in \cite{c12,c13}
assuming the contribution of the nonperturbative gluons and calculating it as the one-pomeron exchange between two
nonperturbative gluons in the collided protons \cite{c13}.
This part was written in the following form \cite{c12}:
\begin{eqnarray}
E(d^3 \sigma/d^3p)_g =
\phi_g(y=0,p_t) \cdot  \sum_{n=2}^{\infty}(n-1) \sigma_n(s) \\
\nonumber
=\phi_g(y=0,p_t) \cdot\left(\sum_{n=1}^\infty n\sigma_n(s)- \sum_{n=1}^\infty \sigma_n(s)\right)= \\
\nonumber
\phi_g(y=0,p_t) \cdot [g(s/s_0)^{\Delta} - \sigma_{nd}]~,
\label{eq:n15}
\end{eqnarray}
where $\sigma_{nd}$ is the non diffractive $pp$ cross section.

Thus, taking into account the quark and gluon contributions we will 
get the following form for the inclusive hadron spectrum:
\begin{eqnarray}
E(d^3 \sigma/d^3p)~=~[\phi_q(y=0,p_t) + \phi_g(y=0,p_t) \cdot\\
\nonumber 
(1 - \sigma_{nd}/g((s/s_0)^{\Delta})]
\cdot g(s/s_0)^{\Delta}
\label{eq:n16}
\end{eqnarray}
%%%%%%%%%%%%%%%%%%%%%%%%%%%%%%%%%%%%%%%%%%%%%%%%%%%%%%%%%%%%%%%%%%%%%%%%%%%%%%%%%%%%%%%%%%%%%%%%%

%%%%%%%%%%%%%%%%%%%%%%%%%%%%%%%%%%%%%%%%%%%%%%%%%%%%%%%%%%%%%%%%%%%%%%%%%%%%%%%%%%%%%%%%%%%%
The question arises, what is the relation of the similarity parameter $\Pi$ to the  
relativistic invariant variables $s,p_t^2$ ? 
This relation can be found from Eqs.(21-24) using $ch(Y)=\sqrt{s}/(2m_0)$.
Then, we have the following form for $\Pi$:
\begin{equation}
\Pi=\left\{\frac{m_{1t}}{2m_0\delta}+\frac{M}{\sqrt{s}\delta}\right\}
\left\{1+\sqrt{1+\frac{M^2-m_1^2}{m_{1t}^2}\delta}\right\}
\label{def:Pi}
\end{equation}
where $\delta=1-4m_0^2/s$; $m_{1t}=\sqrt{p_{t}^2+m_1^2}$ is the transverse mass of the produced hadron $h$.
At large initial energies $\sqrt{s}>>$ 1 GeV the similarity parameter $\Pi$ becomes as following:
\begin{equation}
\Pi=\frac{m_{1t}}{2m_0(1-4m_0^2/s)}
\left\{1+\sqrt{1+\frac{M^2-m_1^2}{m_{1t}^2}(1-4m_0^2/s)}\right\}
\label{def:Picorr}
\end{equation}
For $\pi$-mesons $m_1=\mu_\pi$ is the pion mass and $M=0$; for $K^-$-mesons $m_1=m_K$ is the kaon mass and
$M=m_K$; for $K^+$-mesons $m_1=m_K$ and $M=m_\Lambda - m_0$, $m_\Lambda$ is the mass of the $\Lambda$-baryon.   
For $\pi$-mesons at $p_t^2>>m_1^2$ we have: 
\begin{equation}
\Pi\simeq \frac{m_{1t}}{m_0(1-4m_0^2/s)} 
\label{def:Piappr}
\end{equation}
One can see that in the general case the similarity parameter $\Pi$ 
depends on $p_t^2$ and $s$ and asymptotically at large $s>>4m_0^2$ it depends only on $p_{t\pi}^2$. Let us
stress that the dependence of $\Pi$ on $s$ is crucial at low initial energies only.

\subsection{Quark-gluon dynamics of soft $NN$ interaction and self-similarity function  $\Pi$}
 
The invariant inclusive spectrum can be also presented in the following equivalent form: 
\begin{equation}
\rho_{NN}\equiv E_h\frac{d^3 \sigma_{NN}}{d^3p_h}~=~\frac{1}{\pi}\frac{d\sigma}{dp_t^2dy}\equiv \frac{1}{\pi}\frac{d\sigma}{dm_{1t}^2dy}
\label{def:spypt}
\end{equation} 

Taking into account (\ref{def:spypt}) we can rewrite Eq.(\ref{eq:n16}) in the form:
\begin{eqnarray}
\frac{1}{\pi}\frac{d\sigma_{NN}}{dm_{1t}^2dy}=
[\phi_q(y=0,\Pi) + \phi_g(y=0,\Pi) \cdot(1 -          \\
\nonumber
\sigma_{nd}/g((s/s_0)^{\Delta})]
\cdot g(s/s_0)^{\Delta}~.
\label{eq:n21}
\end{eqnarray}
 The first part of the 
inclusive spectrum (Soft QCD (quarks)) was calculated in \cite{c12,c13} within the QGSM \cite{ABK:1982,ABK:1999}
and then, the function $\phi_q(y=0,\Pi)$ was fitted by the following form \cite{c13,ALM:2015}:
\begin{equation}
\phi_q(y=0,\Pi)~=~A_q exp(-\Pi/C_q)~,
\label{eq:fiqfit}
\end{equation} 
where $A_q=$3.68~(GeV$/$c)$^{-2}$, C$_q=$0.147 for $pp\rightarrow \pi X$ processes at the mid-rapidity and width region of initial energies.

The function $\phi_g(y=0,\Pi)$ related to the second part (Soft QCD (gluons)) of the spectrum, which was calculated in \cite{c13}. 
Then, for the pion production in $pp$ collision at high energies it is fitted by the following form \cite{c13,ALM:2015}:
\begin{equation}
\phi_g(y=0,\Pi)~=~A_g\sqrt{m_{1t}} exp(-\Pi/C_g)~,
\label{eq:figfit}
\end{equation}
where $A_g=$1.7249~(GeV$/$c)$^{-2}$, $C_g$=0.289.     

\begin{figure}[hbtp] 
%\begin{center}
%\begin{tabular}{cc}
\resizebox{1.00\textwidth}{!}{%
\includegraphics{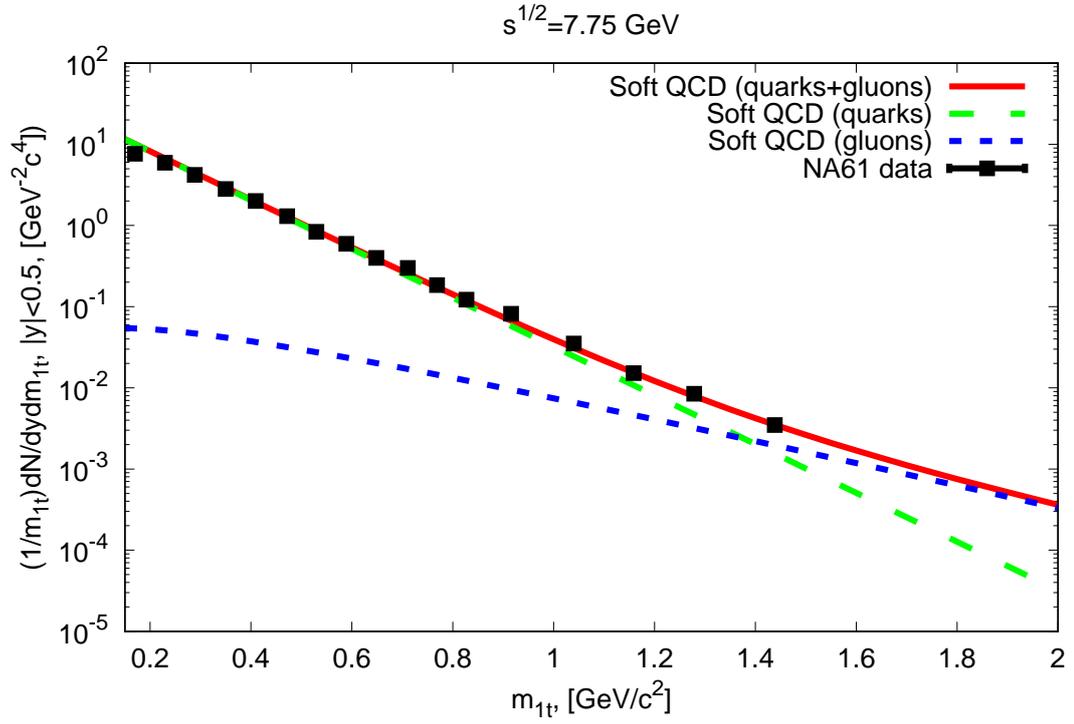}
%Spectrum_pimin_31_crop
}
%\includegraphics[width=0.58\textwidth]{Spectrum_pimin_158}
%\end{tabular}
%\end{center}
\caption{
Results of the calculations of the inclusive cross section of pion 
production in $pp$ collisions as a function of the transverse mass at
$s^{1/2}=$7.75~GeV  or at the initial momentum
$P_{in}=$31~GeV$/c$ in l.s.m. They are compared to the NA61
experimental data \cite{c14}.
} 
\label{fig2}
\end{figure}

\begin{figure}[hbtp] 
%\begin{center}
%\begin{tabular}{cc}
\resizebox{1.00\textwidth}{!}{%
\includegraphics{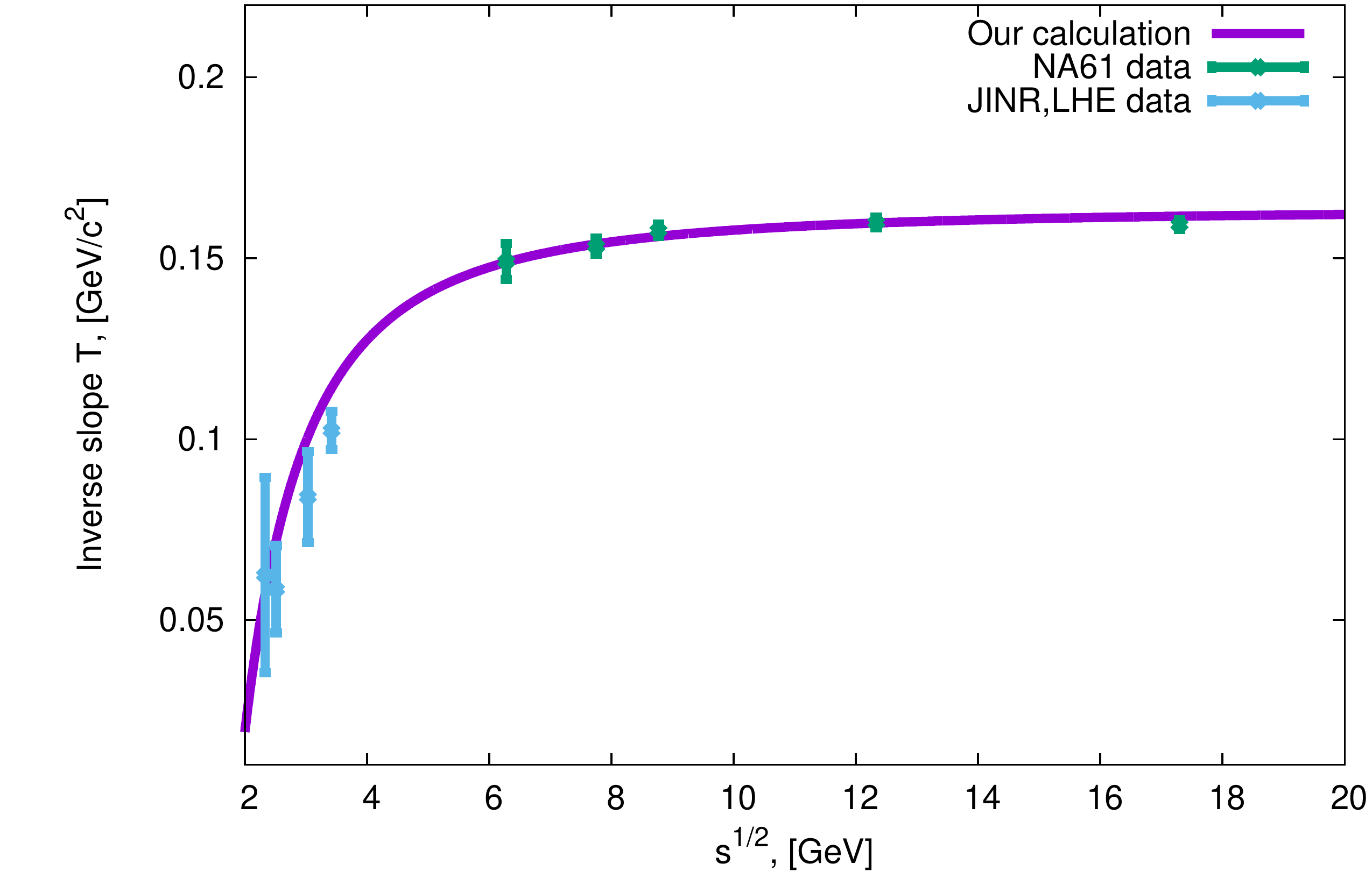} 
}
%%\\[10pt]
%\resizebox{0.55\textwidth}{!}{
%\includegraphics{ATLAS_7T.pdf}
%%spectrum_7T_crop.pdf
%}
 \caption{
The inverse slope parameter $T$ for the reaction $pp->\pi^- X$ calculated using Eq.~\ref{def:Ts}. 
The points at 6 GeV$<\sqrt{s}<$18 GeV are the NA61 data \cite{c14}, the points at 2 GeV $<\sqrt{s}<$ 4 GeV  
are the data extracted from \cite{AJ:2015,AJ:2017}.
%Bottom: results of the calculations of the inclusive cross section of charge hadrons
%produced in $pp$ collisions at the LHC energies as a function of their transverse momentum $p_t$
%at $\sqrt{s}=$7~TeV. The points are the LHC experimental data \cite{CMS,ATLAS}.
%%LHC experimental data  \cite{c14}. 
}  
\label{fig3}
\end{figure}

\begin{figure}[hbtp] 
%\begin{center}
%\begin{tabular}{cc}
\resizebox{1.00\textwidth}{!}{%
\includegraphics{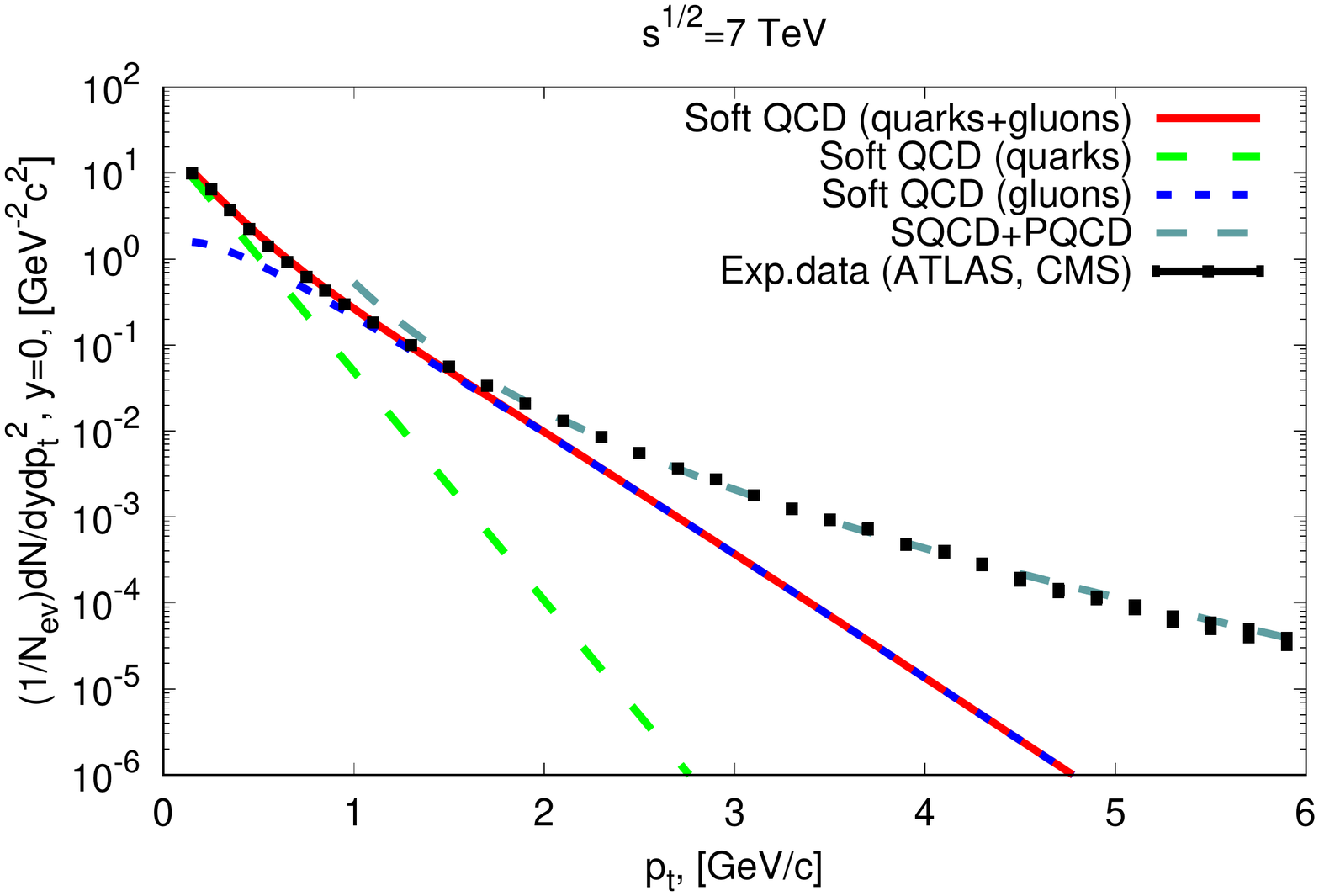}
%Spectrum_pimin_31_crop
}
%\includegraphics[width=0.58\textwidth]{Spectrum_pimin_158}
%\end{tabular}
%\end{center}
\caption{
Results of the calculations of the inclusive cross section of charge hadrons
produced in $pp$ collisions at the LHC energies as a function of their transverse momentum $p_t$
at $\sqrt{s}=$7~TeV. The points are the LHC experimental data \cite{CMS,ATLAS}.
} 
\label{figATL_7T}
\end{figure}

Using (\ref{eq:n21}) we can calculate the inclusive hadron spectrum 
as a function of the transverse mass.                                  
    
    In Fig.~\ref{fig2} the inclusive spectrum
$(1/m_{1t})d\sigma/dm_{1t}dy$ of $\pi^-$-mesons produced in $pp$ collisions at the initial momenta $P_{in}=$ 31 GeV$/$c 
($\sqrt{s}=$7.75 GeV) per nucleon is presented versus their transverse mass $m_{1t}$. 
The similar satisfactory description of the NA61 data at $P_{in}=$ 158 GeV$/$c ($\sqrt{s}=$17.29 GeV) was obtained in our paper \cite{ALM:2015}.
%at the rapidity region 0$<y<$0.2.
Using only the first part of the spectrum $\phi_q(y=0, m_{1t})$, which corresponds to
the quark contribution, the conventional string model, let's call it the SOFT QCD (quarks), one can describe the NA61 data \cite{c14}
rather satisfactorily at  $m_t<$ 1~GeV$/$c$^2$ . This part of the inclusive spectrum 
corresponds to the dashed line in Fig.~\ref{fig2}.
The inclusion of the second part of spectrum due to the contribution of gluons (SOFT QCD (gluons)), the dotted line, allowed 
us to describe all the NA61 data up to $m_{1t}=$ 1.5 Gev$/$c$^2$, see the solid line in Fig.~\ref{fig2} (Soft QCD(quarks+gluons)). 
Actually, at large $\sqrt{s}$ even at the NA61 energies $\Pi\simeq m_{1t}/m_0$ instead of ~(\ref{def:Piappr}).
Generally the pion spectrum $\rho_{NN}(s,m_{1t}\equiv E_h(d^3 \sigma/d^3p_h)$
(ignoring the gluon part) can be presented in the following 
approximated form, which is valid for the NA61 energies and low transverse momenta $p_t<$ 1~GeV$/$c:
\begin{eqnarray}
%\begin{equation}
%\be
\rho_{NN}(s,m_{1t})~\simeq~\phi_q(y=0,\Pi)g(s/s_0)^{\Delta}= g(s/s_0)^{\Delta}\cdot~~~\\ 
\nonumber
A_q exp(-\frac{m_{1t}}{C_qm_0(1-4m_0^2/s)})
%\nonumber
\equiv g(s/s_0)^{\Delta}A_q exp(-\frac{m_{1t}}{T})~, 
\label{def:rhoq_appr}
%\ee
%\end{equation}
\end{eqnarray}
where 
\begin{eqnarray}
T=C_qm_0(1-4m_0^2/s)
\label{def:Ts} 
\end{eqnarray}
is the inverse slope parameter, which is called sometimes as the {\it thermal freeze-out} temperature.
One can see from Eq.(\ref{def:Ts}) that this {\it thermal freeze-out} temperature depends on the initial energy square $s$ in the c.m.s.
of collided protons. That is the direct consequence of the self-similarity approach, which uses the four-momentum velocity formalism. 
This $s$-dependence of $T$ is significant at low initial energies and at $s>>m_0^2$ the inverse slope parameter $T$ becomes independent on $s$.
To describe rather well the NA61 data at larger values of $p_t$, the inclusive pion spectrum should be presented 
by Eq.~(12).
%, which has a more complicated form in comparison to the simple exponential form of 
%Eq.~(\ref{def:rhoq_appr}). 
However, the main contribution to the inelastic total cross section comes from the
first part of Eq.(12), which has the form given by Eq.~(15). 
We have calculated the inverse slope parameter T for the pion production in $pp$ collision as a function of the energy $\sqrt{s}$ given by Eq.(\ref{def:Ts}) 
and presented in Fig.~\ref{fig3}. There is a good agreement with the NA61 data \cite{c14} at 6 GeV $<\sqrt{s}<$18 GeV and the JINR data
\cite{AJ:2015,AJ:2017} at 2 GeV $<\sqrt{s}<$ 4 GeV.
     
%%%%%%%%%%%%%%%%%%%%%%%%%%%%%%%%%%%%%%%%%%%%%%%%%%%%%%%%%%%%%%%%%%%%%%%%%%%%%%%%%%%%%%%%%%%%%%%%
As, for example, in Fig.~\ref{figATL_7T}, 
 we illustrate the satisfactory description of LHC data on inclusive spectrum of charged hadrons (mainly pions and kaons) 
at $\sqrt{s}=$ 7 TeV
by using Eq.(12) and the perturbative QCD (PQCD) within the LO \cite{c12,c13}.  
This spectrum is the sum of inclusive spectra of pions and kaons, therefore it is presented as a function of 
the transverse momentum $p_t$ instead of functions of the transverse mass $m_{1t}$ because the masses of a pion and
kaon are different.    
In addition to the part of spectrum, which corresponds to Eq.(12), see the solid line in this figure, 
we also have included the PQCD calculations, see the dotted line. The PQCD calculations within the LO are 
divergent at low $p_t$, therefore, the dotted line goes up, when $p_t$ decreases. The kinematical region 
about $p_t\simeq$ 1.8-2.2 GeV$/$c can be treated as the matching region of the nonpertubative QCD (soft QCD) and
the pertubative QCD (PQCD). One can see from Fig.~(\ref{figATL_7T}) that it is possible to describe rather well
these inclusive spectrum in the wide region of $p_t$ at the LHC energies matching these two approaches. 
To describe rather well the LHC data on these inclusive $p_t$-spectra at $p_t>$ 2-3 GeV$/$c, the PQCD calculation should be 
included, whose contribution has a shape similar to the power law $p_t$-distribution \cite{CLPST:2014}.
Let us stress that the NA61 data and LHC ones on hadron transverse momentum spectra in $pp$ collisions at the mid-rapidly region
are described within our approach rather satisfactorily with $\chi^2/n.d.f.=$ 0.98 \cite{c15}.  

\section{Nucleus-nucleus collisions in the central rapidity region}
\label{sec:4}

\begin{figure}[hbtp] 
%\begin{center}
%\begin{tabular}{cc}
\resizebox{0.47\textwidth}{!}{%
\includegraphics{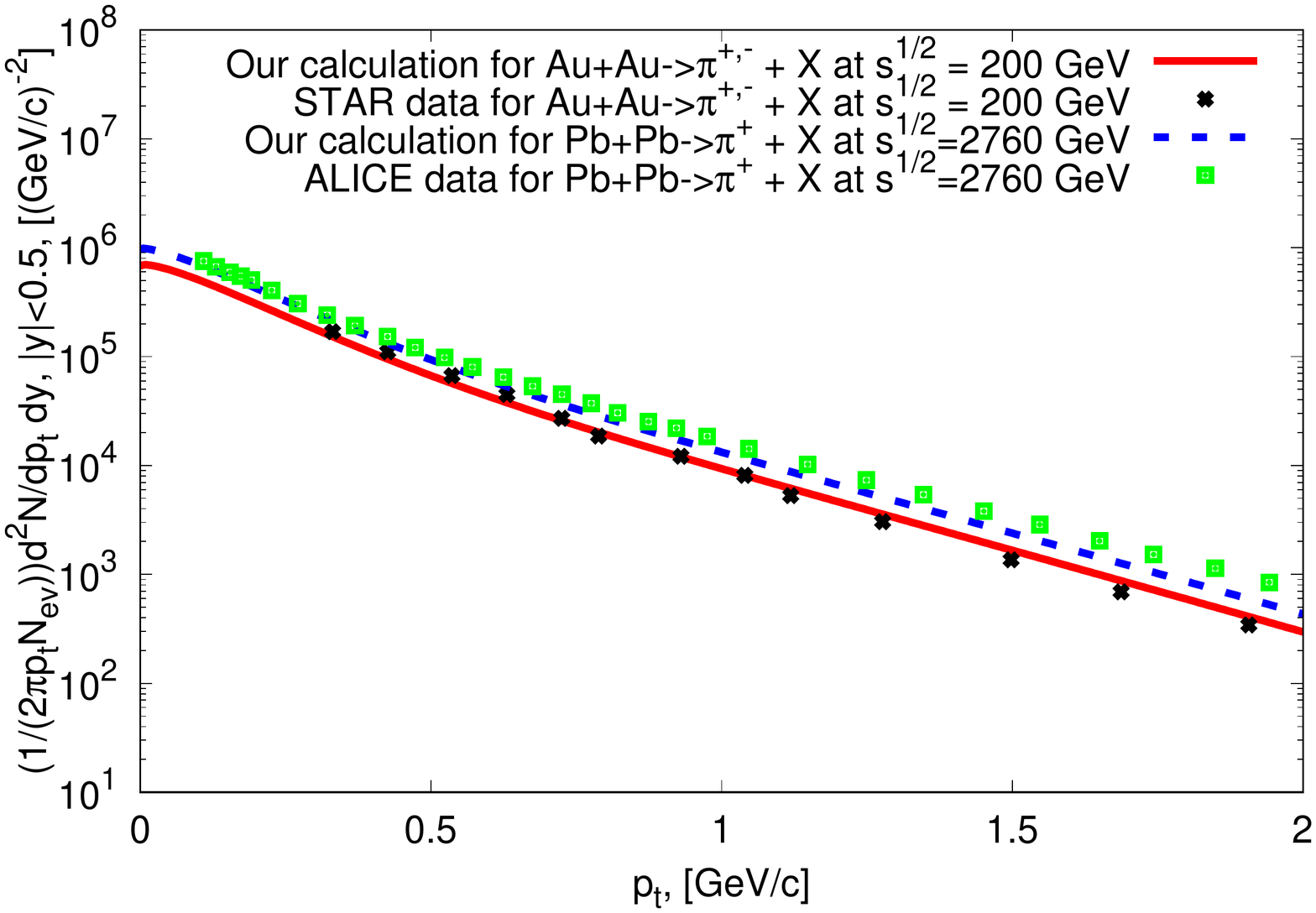}
}
\resizebox{0.47\textwidth}{!}{%
\includegraphics{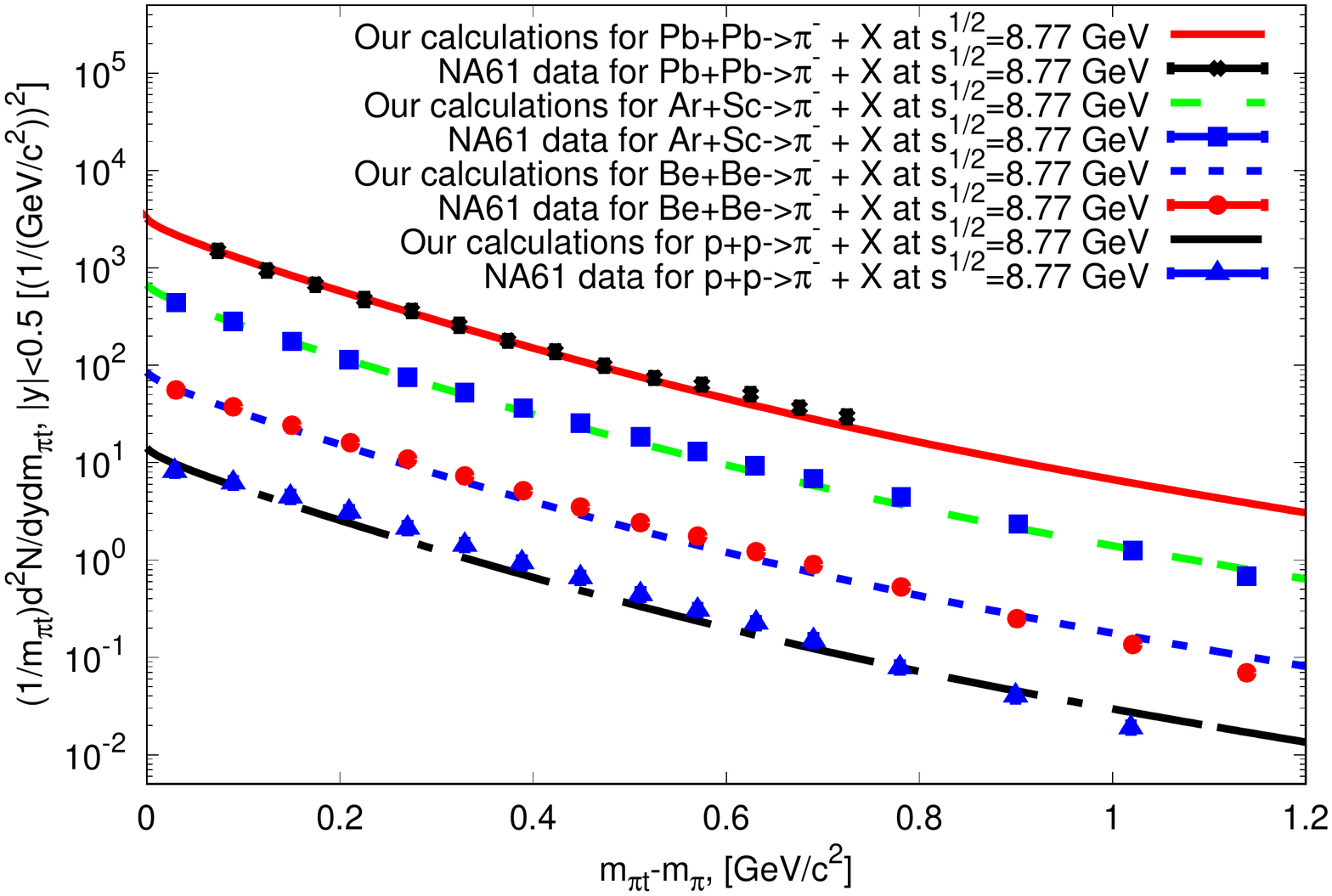}
}
%\end{tabular}
%\end{center}
 \caption{Left: results of our calculations of pion $p_T$-spectra in $AuAu$ and $PbPb$ collisions in the mid-rapidity
region ($|y|<$ 0.5) compared to the STAR \cite{STAR,STAR1} and ALICE \cite{ALICE,ALICE1,ALICE2,ALICE3} data.
Right: Results of our calculations of pion $m_{\pi t}$-spectra in $PbPb,ArSc,BeBe$ and $pp$ collisions at $\sqrt{s}=$ 8.77 GeV
or at the initial momentum per nucleon $P_{in}=$40 GeV$/$c and the mid-rapidity
region compared to the NA61 \cite{NA61_BeBe,NA61_ArSc} data.    
}  
\label{fig4}
\end{figure}

\begin{figure}[hbtp] 
\resizebox{0.47\textwidth}{!}{%
\includegraphics{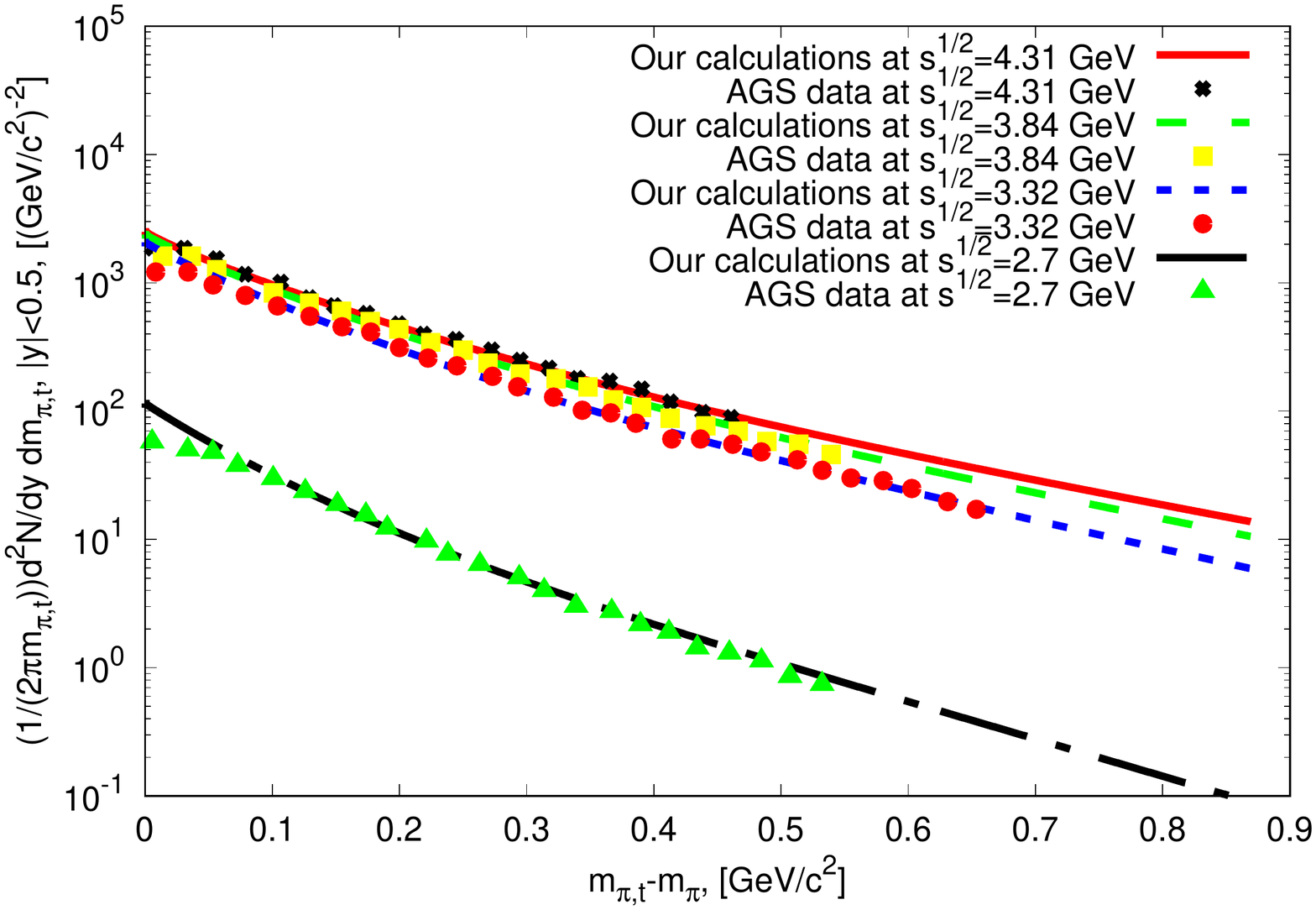} 
}
\resizebox{0.47\textwidth}{!}{%
\includegraphics{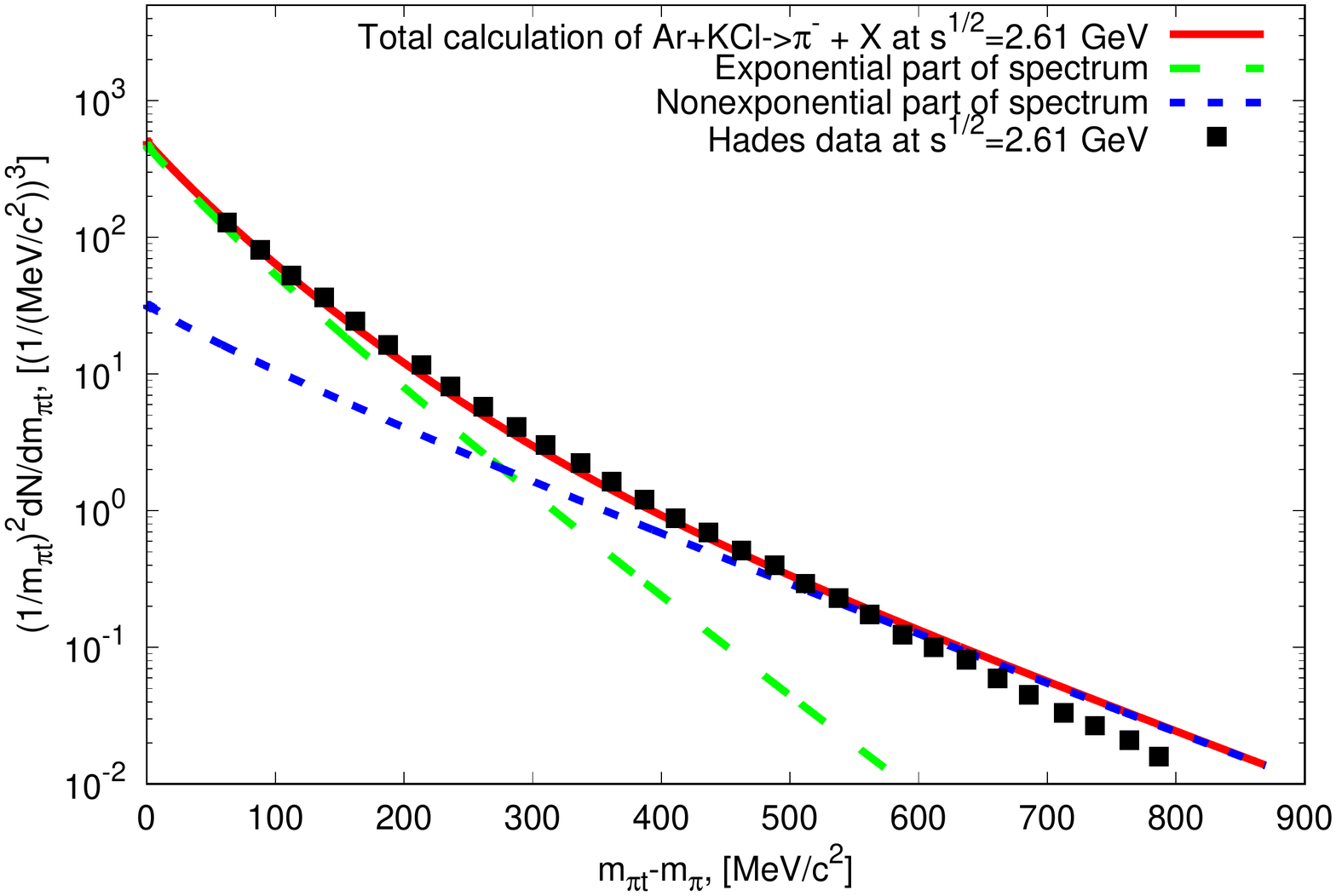}
}
\caption{Left: our calculations of pion $m_{\pi t}$-spectra in $AuAu$ collision in the mid-rapidity region at $\sqrt{s}=$ 4.31, 3.84, 3.32,2.7 GeV
or the initial kinetic energies per nucleon about $E_{kin}=$ 8,6,4,2 GeV respectively. They are compared to the AGS data \cite{AGS}.   
Right: results of our calculations of pion $m_{\pi t}$-spectra in $ArKcl$ collision in the mid-rapidity region at $\sqrt{s}=$ 2.61 GeV or at initial 
kinetic energy per nucleon about 1.75 GeV compared to the HADES data \cite{HADES_ArKcl}; the long dash line corresponds to the exponential part 
of pion spectrum given by Eq.(\ref{eq:fiqfit}) and the short dash curve corresponds to the nonexponential part given by Eq.(\ref{eq:figfit}). 
} 
\label{fig5}
\end{figure}

\begin{figure}[hbtp] 
\resizebox{0.47\textwidth}{!}{%
\includegraphics{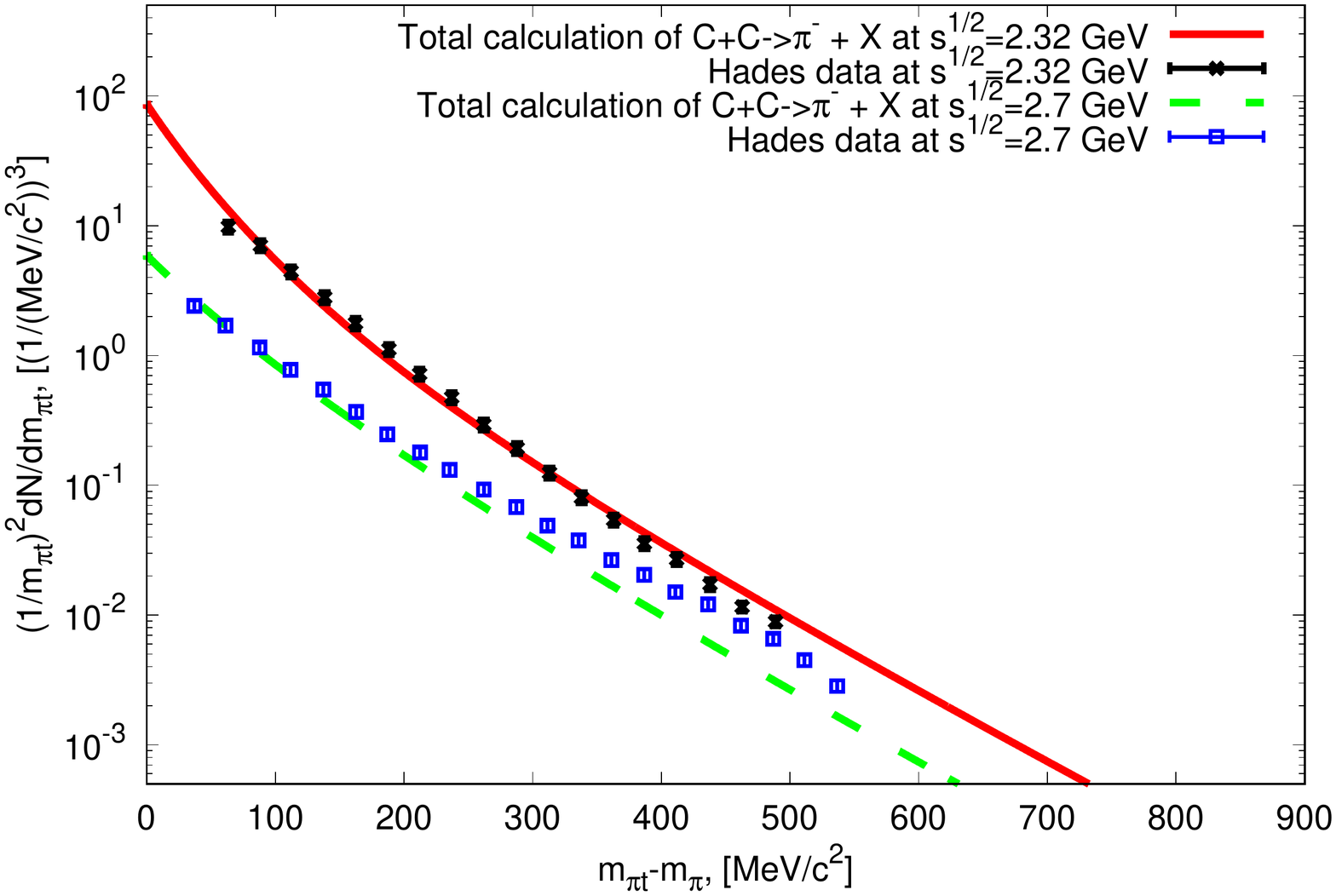} 
}
\resizebox{0.47\textwidth}{!}{%
\includegraphics{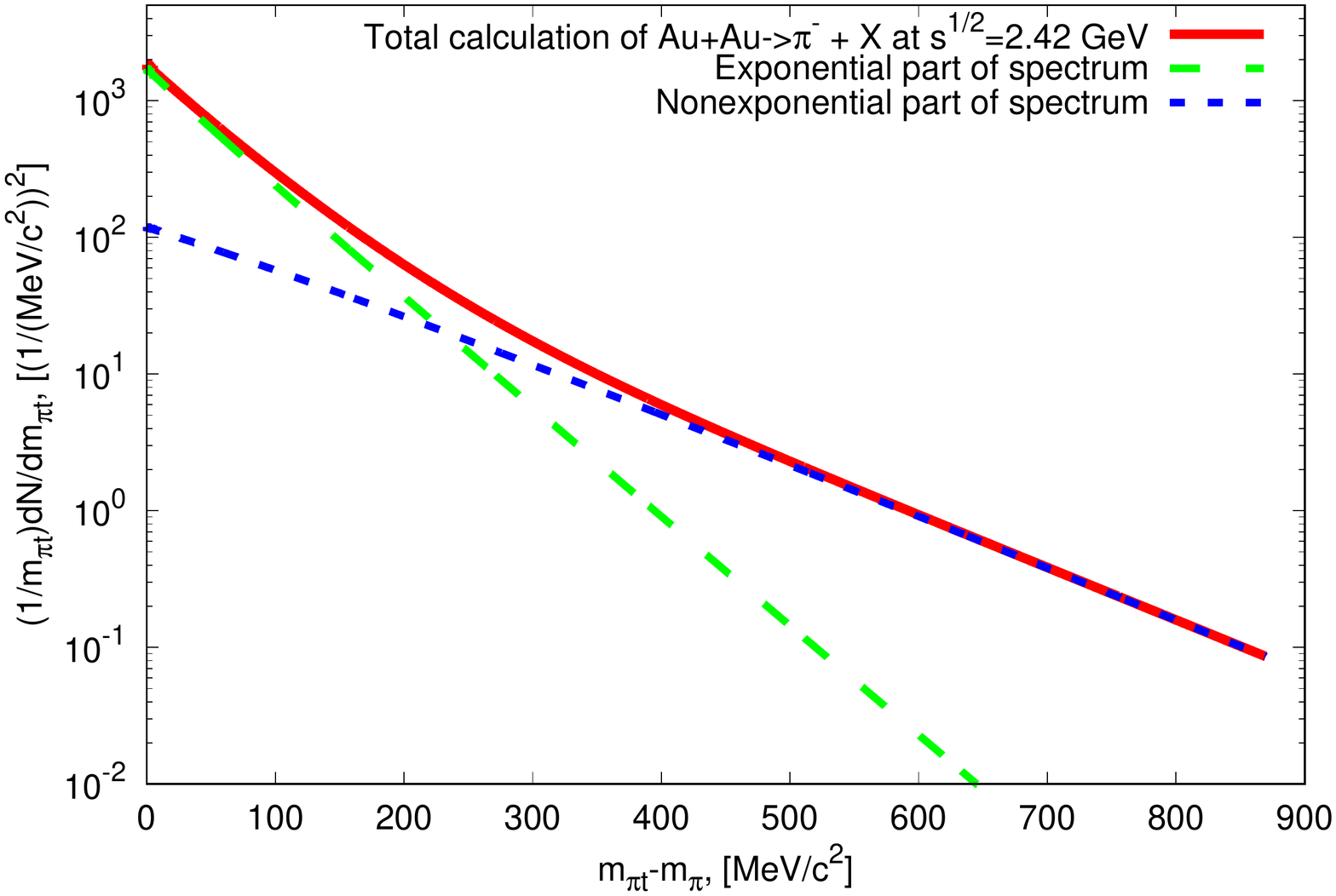}
}
 \caption{Left: results of our calculations of pion $m_{\pi t}$-spectrum in $C^{12}C^{12}$ collision in the mid-rapidity a the initial kinetic energy per
per nucleon about 1 GeV ($\sqrt{s}=$ 2.32 GeV) and 2 GeV ($\sqrt{s}=$ 2.7 GeV) compared to the HADES data \cite{HADES_CC}.
Right: our predictions of pion $m_{\pi t}$-spectrum in $AuAu$ collision in the mid-rapidity region or initial kinetic energy
per nucleon about 1.25 GeV ($\sqrt{s}=$ 2.42 GeV) at the centrality about (0-5)\%; the long dash line corresponds to the exponential part 
of pion spectrum given by Eq.(\ref{eq:fiqfit}) and the short dash curve corresponds to the non exponential part given by Eq.(\ref{eq:figfit}). 
}  
\label{fig6}
\end{figure}

The relativistic invariant inclusive spectrum of hadrons produced in $AA$ collision in the central rapidity region and not large transverse 
momenta can be presented in the following form:
\begin{equation}
E d^3 \sigma_{AA}/dp^3 = C_1 A_I^{\alpha(N_I)} \cdot A_{I\!I}^{\alpha(N_{I\!I})} \cdot \rho_{NN},
\label{def:sigmAA} 
\end{equation}
where $\rho_{NN}$ is the inclusive relativistic invariant pion spectrum in $pp$ collision given by Eq.~(\ref{eq:n21}), 
$\alpha(N)=1/3+N/3$, the $s$-dependent function $N$ is calculated using 
Eqs.~(21-24). The results of our calculations of $p_t$-spectra of pions produced in $AA$ collisions in the central
rapidity region and different high and middle energies compared to different experimental data are presented in Figs.~(\ref{fig4}-\ref{fig6}).
One can see from Fig.~(\ref{fig4}) that our approach is able to get a satisfactory description of the data on pion production in $AuAu$ and $PbPb$ collisions
at the STAR and  LHC energies as well as in $p-p$ collisions, see Figs.~(\ref{fig2},\ref{fig3}) at $p_t\leq$ 1.2 GeV/c. 
In principle, there can be another theoretical interpretation of multiple hadron production in heavy-ion collisions at high energies 
based, for example, on the stationary thermal model \cite{Heinz1:1993,Heinz2:1993}.
At middle initial energies about 1 GeV-8 GeV our calculations of pion $p_t$-spectra in heavy ion collisions, namely $AuAu$, $ArKCl$, at the central 
rapidities result in a more or less satisfactory description of the data shape at low $p_t<$ 0.5-0.6 GeV/c, as it also can be seen in 
Figs.~(\ref{fig4}-\ref{fig6}). 
By calculation of all the pion $p_t$-spectra in $AA$ collisions at mid-rapidity we used the same form of $\rho_{NN}$ applied to the satisfactory description
of the NA61 and  LHC data on hadron production in $pp$ collisions, see Figs.~(\ref{fig4}-\ref{fig6}). The energy dependence of these spectra is given 
by the term $g(s/s_0)^\Delta$ for the 
quark contribution and the term $(g(s/s_0)^\Delta-\sigma_{nd})$ for the gluon contribution to $\rho_{NN}$. The non diffractive 
cross section $\sigma_{nd}$, as a difference between the total $pp$ cross section $\sigma_{tot}$ and the elastic ($\sigma_{el}$) and the diffractive ($\sigma_{dif}$)
cross sections at high energies is taken from the experimental data. At middle energies about several GeV there are very poor data on the diffractive cross section
$\sigma_{dif}$. Therefore, at $\sqrt{s}$ about a few GeV our calculations are not so precise, as at high energies. As one can see from Fig.(\ref{fig4})
the $m_{\pi t}$-pion spectra in heavy ion collisions, as $Au+Au, Pb+Pb$, at high energies
and the mid-rapidity are described rather satisfactorily within the proposed 
approach at $m_{\pi t}<$ 0.7 GeV$/$c$^2$ with $\chi^2/n.d.f.=$0.98.  

%{\bf
A small deviation of our calculations from the HADES data less than 10 \% can be seen in the $m_{\pi t}$-spectra   
of the pions produced in $Ar+KCl$ collision at the initial kinetic energies per nucleon about 1.75 GeV ($\sqrt{s}=$ 2.61 GeV) and 
$m_{\pi t}<$ 0.6 GeV$/$c$^2$ presented in Fig.~\ref{fig5} (right).  It is illustrated by the Table 1 presented in the 
Appendix. Approximately the same deviation, as in the Table 1, is for the pion production in $Au+Au$ collision also at 
$m_{\pi t}<$ 0.6 GeV$/$c$^2$, see Fig.~\ref{fig5} (right).  
However, at $m_{\pi t}>$ 0.6 GeV$/$c$^2$ this deviation can be about 20\%-70\%, as it is seen from the Table 1.

Contrary to this the rather big deviation of our calculations from the HADES data is seen in the pion production in $C^{12}+C^{12}$ 
collision at the initial kinetic energy per nucleon about 2 GeV presented in Fig.~\ref{fig6} (left) especially at $m_{\pi t}>$ 0.3 GeV$/$c$^2$.
However, the AGS data for $Au+Au\rightarrow\pi + X$ reaction at the same energy per nucleon are described more better, as it is seen 
in Fig.~\ref{fig5} (left).   
%}

A big deviation of the theory from the data
is not seen in $p_t$-spectra of the pions produced in heavy ion collisions, for example, $AuAu, PbPb$. It can be due to the difference between nuclear 
density $\rho_N(k)$ distribution in heavy nuclei and light nuclei. In the heavy nucleus $\rho_N(k)$ as function of the internal nucleon momentum $k$ is 
more flat compared to the nucleon distribution in the light nucleus, therefore the pion production in heavy ion collision can be less sensitive to the nuclear 
structure compared to the same pion production in light nucleus-nucleus collisions.      

In Fig.~\ref{fig6} (right) we present the prediction of pion $m_{\pi t}$-spectrum in $AuAu$ collision in the mid-rapidity region and
centrality about (0-5)\% for the HADES experiment at initial kinetic energy per nucleon about 1.25 GeV ($\sqrt{s}=$ 2.42 GeV).        

\section{Conclusion}

%\label{sec:5}
\begin{sloppypar} 
 
The inclusive hadron spectrum in the space of four-velocities is presented within the self-similarity
approach as a function of the similarity parameter $\Pi$.
The use of the self-similarity approach allows us to describe the ratio of the total yields
of protons to the anti-protons produced in $AA$ collisions as a function of the energy in the mid-rapidity region
and a wide energy range from 10~GeV to a few TeV \cite{ALM:2015}. 

We have shown that the energy dependence of the similarity parameter $\Pi$ included within this approach is very significant at low energies,
namely at $\sqrt{s}<$ 6~GeV, and rather well reproduces the experimental data on the inverse slope or the {\it thermal freeze-out}
temperature of the inclusive spectrum of the hadrons produced in $pp$ collisions. The parameter $\Pi$ increases and saturates when $\sqrt{s}$ grows. 
This is very significant for a theoretical interpretation 
of the future experimental data planned to get at FAIR,~CBM (Darmstadt, Germany), RHIC (BNL, Brookhaven, USA) and NICA (Dubna, Russia) projects. That is 
an advantage of the self-similarity approach compared to other theoretical models.

However, we have also shown that the $s$ dependence of $\Pi$ is not enough to describe
the inclusive spectra of the hadrons produced in the mid-rapidity region, for example, in $pp$ collisions in the wide region of
initial energy, especially at the LHC energies. Therefore, we have modified the self-similarity approach using the quark-gluon string model 
(QGSM) \cite{ABK:1982,ABK:1999} and \cite{c12,c13} including the contribution of nonperturbative gluons, which are very
significant to describe the experimental data on inclusive hadron spectra in the mid-rapidity region at the transverse momenta 
$p_t$ up to 2-3~GeV$/$c \cite{c12,c13}. Moreover, the gluon density obtained in \cite{c13}, whose parameters were found from the best 
description of the LHC data and also allowed us to describe the HERA data on the proton structure functions \cite{c15}. To describe the data 
in the mid-rapidity region and values of $p_t$ up to 2-3~GeV$/$c, we have modified the simple exponential form of the spectrum, as a function of $\Pi$,  
and presented it in two parts due to the contribution of quarks and gluons, each of them has a different energy dependence. This energy dependence
was obtained in \cite{c12} by using the Regge approach valid for soft hadron-nucleon processes. 
To extend the application of the offered approach to analyze these inclusive $p_t$-spectra at large hadron transverse momenta, we have to 
take the PQCD calculations into account.

This approach is applied to the analysis of pion production in $pp$, $AA$ and $pA$ collisions in the mid-rapidity region.
We have shown the self-consistent satisfactory description of the data on $p_t$-spectra of the pions in these interactions 
in a wide region of initial energies and not large transverse momenta of pions.
The approach suggested in this paper results in a more or less well description of these spectra for heavy-ion collisions. 
However, it can not be applied to the analysis of hadron production in light nucleus-nucleus collisions, especially, at middle 
energies because the production mechanism is very sensitive to the nuclear structure, which is different for heavy and light nuclei.      
\end{sloppypar}

{\bf Acknowledgements.}

\begin{sloppypar} 
We are very grateful to A.P. Jerusalimov, A.A. Baldin, A.V. Belayev, A.Yu. Troyan for giving us the experimental data obtained at the JINR.
We thank M. Gumberidze, R. Holzmann, G. Kornakov, A. Rustamov, J. Stroth for extremely helpful discussions and very productive collaboration. 
We are also grateful to P. Braun-Munzenger, T. Galatyuk, V.P. Ladygin, M. Lorenz, V. Pechenov, B. Ramstein, P. Salabura for helpful discussions.  
\end{sloppypar}

\vspace{0.5cm}

{\bf APPENDIX}

\vspace{0.3cm}

In this paragraph we present the analytical form of the self-similarity function 
$\Pi$.
Equation (2) can be written as follows:
\begin{equation}
 N_I \cdot N_{II} - \Phi_I \cdot N_I - \Phi_{II} \cdot N_{II} = \Phi_M \quad ,
\label{def:eq1}
\end{equation}
where relativistic invariant dimensionless values have been introduced:

$$ \Phi_I = [(m_1/m_0) \cdot (u_Iu_1) + M/m_0]/[(u_Iu_{II})- 1]$$
$$ \Phi_{II} = [(m_1/m_0) \cdot (u_{II}u_1) + M/m_0]/[(u_Iu_{II})- 1]$$
$$ \Phi_M = (M^2 - m^2_1)/[2m_0^2((u_Iu_{II})- 1)].$$
     Equation (18) can be written as follows:
\begin{equation}
[(N_I/  \Phi_{II})- 1] \cdot [(N_{II}/ \Phi_I) - 1]=1 + 
[\Phi_M/(\Phi_I \cdot  \Phi_{II})].                           
\label{def:eq2n}
\end{equation}

Minimum $\Phi$ is found from the following:
\begin{equation}             
d \Pi /dN_I = 0 , \qquad  d\Pi/dN_{II} = 0.
\label{def:eq3}
\end{equation}

     Let us introduce the intermediate variables:
$$  F_I = [(N_I/ \Phi_{II}) - 1],  \qquad  F_{II}=[(N_{II}/  \Phi_I)  - 1].$$

From the above we obtain: $F_I \cdot F_{II} = 1+ \Phi_M/(\Phi_I \cdot \Phi_{II}).$

Then, (20) is also equal to 0 as
$$ d \Pi/dF_I = 0,  \qquad   d \Pi/dF_{II} = 0.$$
From (3) we can obtain:
$$ 4\Pi^2 = N_I^2 + N_{II}^2 + 2N_I \cdot N_{II} \cdot(u_Iu_{II}),$$
$$4\Pi^2 = (F_I + 1)^2 \Phi_{II}^2 + (F_{II} + 1)^2 \Phi_I^2 +$$
$$2 \Phi_I \cdot \Phi_{II} (F_I +1) \cdot (F_{II}+ 1) \cdot(u_Iu_{II})
F_{II}  = \alpha /F_I. $$
     The condition of the minimum $d(4\Pi^2)/dF_1 = 0$ gives the 
equation for $F_I$: 
$$F_I^4 + F_I^3 - (\Phi_I/ \Phi_{II})^2  \cdot (\alpha^2 + \alpha F_I) +\\
(u_Iu_{II}) \cdot $$
$$ (\Phi_I/ \Phi_{II}) \cdot (F_I^3 - \alpha  F_I) = 0$$
or
$$F_I^4 + F_I^3 [1 + (u_Iu_{II})/z] -  (\alpha/z) \cdot F_I \cdot [(u_Iu_{II}) +$$
$$(1/z)] - \alpha^2/z^2 = 0,$$
where $z = \Phi_{II}/ \Phi_I$.
Changing I to II we should replace  $z \to (1/z),  F_I \to (\alpha/F_{II}).$
$$(\alpha/F_{II})^4 + (\alpha/F_{II})^3 [1 + (u_Iu_{II})z] -\\
\alpha z(\alpha/F_{II})\cdot $$
$$[(u_Iu_{II}) + z]-\alpha^2z^2=0$$
or
$$F_{II}^4 + F_{II}^3[1 + (u_Iu_{II})z] - z\alpha \cdot F_{II}\cdot [z +
(u_Iu_{II})] - \alpha^2z^2 = 0.$$
Thus, at $z=1 \to F_I=F_{II},$ $ \Phi_I =\Phi_{II} =\Phi.$\\
Since $F_I = F_{II}$, then $(N_I/ \Phi- 1)=(N_{II}/ \Phi- 1)$
and $N_I = N_{II}$.\\
$ F^2 =\alpha$  and  $F_I= F_{II} =\alpha^{1/2}=[1 + (\Phi_M/ \Phi^2)]^{1/2}.$
\begin{eqnarray}
N_I =N_{II}= N=(1+F)\Phi = \\
\nonumber
\{ 1 + [1 + (\Phi_M/ \Phi^2)]^{1/2} \} \Phi
\end{eqnarray}
\begin{eqnarray}
\Pi = 1/2 [2N^2 + 2N^2(u_Iu_{II})]^{1/2}=\\
\nonumber
N/ \sqrt{2})[1 + (u_Iu_{II})]=N \cdot ch Y. 
\end{eqnarray}
Note that $(u_Iu_{II}) = ch2Y$,
$(u_Iu_1)=(m_t/m_1) \cdot ch(-Y-y) =(m_t/m_1) \cdot ch(Y+y)$
and
$(u_{II}u_1) = (m_t/m_1) \cdot ch(Y-y)$.
Here $m_t$ is the transverse mass of the particle 1,  
$m_t = (m_1^2 + p_t^2)^{1/2}$,
$Y$ - rapidity of interacting nuclei, $y$ - particle 1 rapidity.
At $y = 0$ (in the central rapidity region) we obtainthe following:
$$(u_Iu_1)=(u_{II}u_1)=(m_{1t}/m_1) \cdot chY, \quad m_{1t} =(m_1^2 + p_t^2)^{1/2}$$
\begin{eqnarray}
\Phi= (1/m_0) \cdot (m_{1t}chY + M) \cdot [1/(2sh^2Y)]
\end{eqnarray}
\begin{eqnarray}
\Phi_M= (M^2 - m_1^2)/(4m_0^2 sh^2 Y)
\end{eqnarray}
\begin{table}[ht]
  \setlength{\tabcolsep}{8pt}
  \centering
  \begin{tabular}{|r|r|r|} \hline 
%   $MeV/c^2$                   
  \multicolumn{1}{|c}{$m_{\pi t}-m_\pi$} & \multicolumn{1}{|c|}{Theory} & \multicolumn{1}{c|}{Exp. data}  \\
  \multicolumn{1}{|c}{MeV$/$c$^2$}  & \multicolumn{1}{|c|}{(MeV$/$c$^2$)$^{-3}$} & \multicolumn{1}{c|}{(MeV$/$c$^2$)$^{-3}$} \\ 
\hline  
    6.259e+01 & 1.297e+02 &   1.278e+02 \\   
    8.819e+01 & 7.958e+01 &   8.066e+01 \\ 
    1.124e+02 & 4.949e+01 &   5.248e+01 \\ 
    1.380e+02 & 3.270e+01 &   3.631e+01 \\  
    1.622e+02 & 2.173e+01 &   2.436e+01 \\  
    1.878e+02 & 1.456e+01 &   1.634e+01 \\  
    2.134e+02 & 9.870e+00 &   1.166e+01 \\ 
    2.361e+02 & 7.045e+00 &   8.066e+00 \\
    2.617e+02 & 4.954e+00 &   5.754e+00 \\ 
    2.873e+02 & 3.506e+00 &   4.105e+00 \\
    3.101e+02 & 2.622e+00 &   3.020e+00 \\
    3.371e+02 & 1.881e+00 &   2.222e+00 \\
    3.613e+02 & 1.421e+00 &   1.634e+00 \\
    3.869e+02 & 1.067e+00 &   1.202e+00 \\
    4.111e+02 & 8.215e--01  &   8.844e-01 \\
    4.367e+02 & 6.283e-01  &   6.918e-01 \\
    4.623e+02 & 4.844e-01  &   5.089e-01 \\
    4.879e+02 & 3.761e-01  &   3.981e-01 \\
    5.121e+02 & 2.992e-01  &   2.929e-01 \\
    5.377e+02 & 2.349e-01  &   2.291e-01 \\
    5.619e+02 & 1.886e-01  &   1.738e-01 \\
    5.875e+02 & 1.499e-01  &   1.240e-01 \\
    6.117e+02 & 1.211e-01  &   1.000e-01 \\
    6.373e+02 & 9.642e-02  &   8.066e-02 \\
    6.615e+02 & 7.826e-02  &   5.934e-02 \\
    6.856e+02 & 6.363e-02  &   4.501e-02 \\
    7.127e+02 & 5.053e-02  &   3.311e-02 \\
    7.368e+02 & 4.120e-02  &   2.671e-02 \\
    7.639e+02 & 3.294e-02  &   2.089e-02 \\
    7.866e+02 & 2.725e-02  &   1.585e-02 \\
\hline
\end{tabular}
\caption{ Our calculations of the pion spectrum $(1/m_{\pi t})^2 dN/dm_{\pi t}$ 
 presented in Fig.~\ref{fig5} as a function of $m_{\pi t}-m_\pi$ (MeV$/$c$^2$) in $ArKCl$ collision at the mid-rapidity region 
and initial kinetic energy per nucleon about 1.75 GeV ($\sqrt{s}=$ 2.61 GeV), second column, compared to the HADES data 
\cite{HADES_ArKcl} (third column). 
}
\label{tab:cl}
\end{table}
%%%%%%%%%%%%%%%%%%%%%%%%%%%%%%%%%%%%%%%%%%%%%%%%%%%%%%%%%%%%%%%%%%%%%%%%%%%%


\begin{thebibliography}{00}
\bibitem{Fermi:1950}
{E.~Fermi}, Phys. Rev. \textbf{92}, (1953) 452.
\bibitem{Pomeran:1951}
{I. Ya.~Pomeranchuk}, Izv. Dokl. Akad. Nauk Ser.Fiz. \textbf{78} (1951) 889.
\bibitem{Landau:1953}
{L.D.~Landau}, Izv. Akad. Nauk Ser. Fiz. \textbf{17} (1953) 51.
\bibitem{Hagedorn:1965} 
{R.~Hagedorn}, Supplemento al Nuovo Cimento \textbf{3}, 147 (1965).
\bibitem{PBM:2013}
P.~Braun-Munzinger, K.~Redlich, J.~Stachel, Nucl.Phys. \textbf{A904} (2013) 535c.
\bibitem{Chatt:2015}
S.~Chatterjee et al., Advances in High Energy Physics, vol.2015, ID 349013. 
\bibitem{Bugaev:2018}
K.A.~Bugaev et al., Nucl.Phys. \textbf{A970} (2018) 133.
\bibitem{Br-Munz}
{A.~Andronic, P.~Braun-Munzinger, J.~Stachel}, Nucl. Phys. \textbf{A772} (2006) 167;
arXiv:0511071 [nucl-th].
\bibitem{c3}  
{A.M.~Baldin, L.A.~Didenko}, Fortsch.Phys. \textbf{38} (1990) 261.
\bibitem{c4}  
{A.M.~Baldin, A.I.~Malakhov, and A. N.~Sissakian}, Phys. Part. Nucl. \textbf{29} (Suppl. 1)
(2001), 4.
\bibitem{ALM:2015}
{D.A.~Artemenkov, G.I.~Lykasov, A.I.~Malakhov},
Int.J.Mod.Phys. \textbf{A30} (2015) 1550127.
\bibitem{a1}
{W.~Heisenberg}, Physik und Philosophie, Frankfurt am Main, 1959.
\bibitem{a2} A.M.~Baldin and A.A.~Baldin. Physics of Particles and Nuclei \textbf{29}, 1998,
232.
\bibitem{c5}  {A.M.~Baldin, A.I.~Malakhov}, JINR Rapid Communications 1 [87]-98 (1998) 5.
%\bibitem{c6} A. M. Baldin, A. A. Baldin. Phys. Particles and Nuclei, 29 (3), (1998) 232.
\bibitem{c7} {A.~Tawfik}, Nuclear Physics \textbf{A859}, (2011) 63.
\bibitem{c8}  {\tt http://hepdata.cedar.ac.uk/view/p7907.}
\bibitem{c9} R.~Klingenberg \textit{et al.},  Nuclear Physics \textbf{A610}, (1996) 306c.
%%%%%%%%%%%%%%%%%%%%%%%%%%%%%%%%%%%%%%%%%%%%%%%%%%%%%
\bibitem{c12} {V.A.~Bednyakov, A.A.~Grinyuk, G.I.~Lykasov, M.~Pogosyan};
 Int.J.Mod.Phys. A27 (2012) 1250042. 
\bibitem{ABK:1982} {A.B.~Kaidalov}, Z.Phys. \textbf{C12}, (1982) 63.
\bibitem{ABK:1999} {A.B. Kaidalov}, Surveys High Energy Phys. \textbf{13}, (1999) 265.
\bibitem{c10} {V.~Abramovsky, V. N.~Gribov and O.~Konchelli}, 
Sov.J.Nucl.Phys. \textbf{18}, (1973) 308.
\bibitem{c11} {K.A.~Ter-Martirosyan}, Sov.J.Nucl.Phys., \textbf{44}, (1986) 817
\bibitem{c13} {A.A.~Grinyuk, G.I.~Lykasov, A.V.~Lipatov, N.P.~Zotov}, 
Phys.Rev. \textbf{D87}, (2013) 074017.
\bibitem{c14} {A.A.~Abgrall \textit{et al}, Eur.Phys.J., \textbf{C74}, (2014) 2794.
%\bibitem{Abdiv}
%A.~Abdivaliev, et al., JINR-P1-82-507 (1982). 
\bibitem{AJ:2015}
 A.P.~Jerusalimov \textit{et al.}, Eur.Phys.J. \textbf{A51}, 83 (2015).
\bibitem{AJ:2017}
A.P.~Jerusalimov, \textit{et al.}, EPJ Web Conf. \textbf{138},  07008, (2017).
\bibitem{CMS} V.~Khachatryan, \textit{et al.}, (CMS Collaboration), Phys. Rev. Lett. \textbf{105}, (2010) 022002.
\bibitem{ATLAS} G.~Aad, \textit{et al.}, (ATLAS Collaboration), 
New J. Phys. {\textbf{13}}, (2011) 053033.
\bibitem{CLPST:2014}	
{J.~Cleymans, G.I.~Lykasov, A.S.~Parvan, A.S.~Sorin, O.V.~Teryaev},
Phys.Lett. \textbf{B723}, (2013) 351; arXiv:1302.1970 [hep-ph].
\bibitem{Heinz1:1993} {E.~Schnedermann, U.~Heinz}, Phys.ReV. \textbf{C47}, 1738 (1993).
\bibitem{Heinz2:1993} {E.~Schnedermann, J.~Solfrank, U.~Heinz}, Phys.Rev. \textbf{C48}, 2462 (1993).
\bibitem{STAR} B.I.~Abelev \textit{et al.}, (STAR  Collaboration), Phys. Rev. \textbf{C75}, 064901 (2007).
\bibitem{STAR1} B.I.~Abelev \textit{et al.}, (STAR  Collaboration),
Phys.Rev.Lett., \textbf{97}, 152301 (2006).
%\bibitem{PHENIX} A. Adare et al. (PHENIX Collaboration), Phys. Rev. C {\textbf 83}, 052004, (2010); 
%Phys. Rev. C {\textbf{83}}, 064903 (2011).
\bibitem{ALICE} K.~Aamodt, \textit{et al.}, (ALICE Collaboration), 
Eur. Phys. J. C \textbf{71}} 1655 (2011).
\bibitem{ALICE1} K.~Aamodt, \textit{et al.}, (ALICE Collaboration),
Phys. Lett. \textbf{B693}, 53 (2010).
\bibitem{ALICE2} {K.~Aamodt}, \textit{et al.}, (ALICE Collaboration), 
Phys. Rev. \textbf{D82}, 052001 (2010).
\bibitem{ALICE3} {K.~Aamodt}, \textit{et al.}, (ALICE Collaboration), 
Phys. Rev. \textbf{C88}, 044910 (2013).
\bibitem{NA61_BeBe} {E.~Kaptur}, PoS \textbf{CPOD2014}, (2015) 053
\bibitem{NA61_ArSc} {M.~Lewicki}, arXiv:1612.01334 [hep-ex].
\bibitem{AGS} J.L.~Klay, \textit{et all}, (AGS Collaboration), Phys.Rev. \textbf{C68}, (2003) 054905
\bibitem{HADES_ArKcl} P.~Tlusty, \textit{et all.}, (HADES Collaboration), arXiv:0906.2309 [nucl-ex].
\bibitem{HADES_CC} G.~Agakishev, \textit{et all.},(HADES Collaboration), Eur.Phys.J., \textbf{A40}, (2009) 45.
\bibitem{c15} {A.V.~Lipatov, G.I.~Lykasov, N.P.~Zotov}, Phys.Rev. \textbf{D89}, (2014) 014001.

\end{thebibliography}
\end{document}